\PassOptionsToPackage{table,svgnames,dvipsnames}{xcolor}
\documentclass[journal, hideappendix]{vgtc} 

\onlineid{1207}

\vgtccategory{Research}

\vgtcpapertype{Theoretical \& Empirical}

\title{Qualitative Study for LLM-assisted Design Study Process: Strategies, Challenges, and Roles}

\author{%
  \authororcid{Shaolun Ruan}{0000-0002-6163-9786},
  \authororcid{Rui Sheng}{0000-0001-9321-6756},
  \authororcid{Xiaolin Wen}{0000-0002-8562-7640},
  \authororcid{Jiachen Wang}{0000-0001-9630-9958},
  \authororcid{Tianyi Zhang}{0009-0009-1318-3655},\\
  \authororcid{Yong Wang}{0000-0002-0092-0793},
  \authororcid{Tim Dwyer}{0000-0002-9076-9571},
  and \authororcid{Jiannan Li}{0000-0001-8409-4910}
}

\authorfooter{
  \item
  	S. Ruan, T. Zhang, and J. Li are with Singapore Management University.
  	E-mail: \{slruan.2021, tianyizhang.2023\}@phdcs.smu.edu.sg and jiannanli@smu.edu.sg. S. Ruan is also with Monash University.
  \item
  	R. Sheng is with the Hong Kong
University of Science and Technology.
E-mail: rshengac@connect.ust.hk.

  \item X. Wen and Y. Wang are with Nanyang Technological University. 
  E-mail: xiaolin004@e.ntu.edu.sg and yong-wang@ntu.edu.sg.

    \item J. Wang is with Zhejiang University.
  E-mail: wangjiachen@zju.edu.cn.
  
    \item T. Dwyer is with Monash University.
  E-mail: tim.dwyer@monash.edu.

  \item J. Li and Y. Wang are the co-corresponding authors.
}

\abstract{%
Design studies aim to develop visualization solutions for real-world problems across various application domains.
Recently, the emergence of large language models (LLMs) has introduced new opportunities to enhance the design study process,
providing capabilities such as creative problem-solving, data handling, and insightful analysis. 
However, despite their growing popularity, there remains a lack of systematic understanding of how LLMs can effectively assist researchers in visualization-specific design studies. 
In this paper, we conducted a multi-stage qualitative study to fill this gap, which involved 30 design study researchers from diverse backgrounds and expertise levels. 
Through in-depth interviews and carefully-designed questionnaires, we investigated strategies for utilizing LLMs, the challenges encountered, and the practices used to overcome them. 
We further compiled the roles that LLMs can play across different stages of the design study process. 
Our findings highlight practical implications to inform visualization practitioners,
and also provide a framework for leveraging LLMs to facilitate the design study process in visualization research.
}


\teaser{
  \centering
  \includegraphics[width=\linewidth, alt={A view of a city with buildings peeking out of the clouds.}]{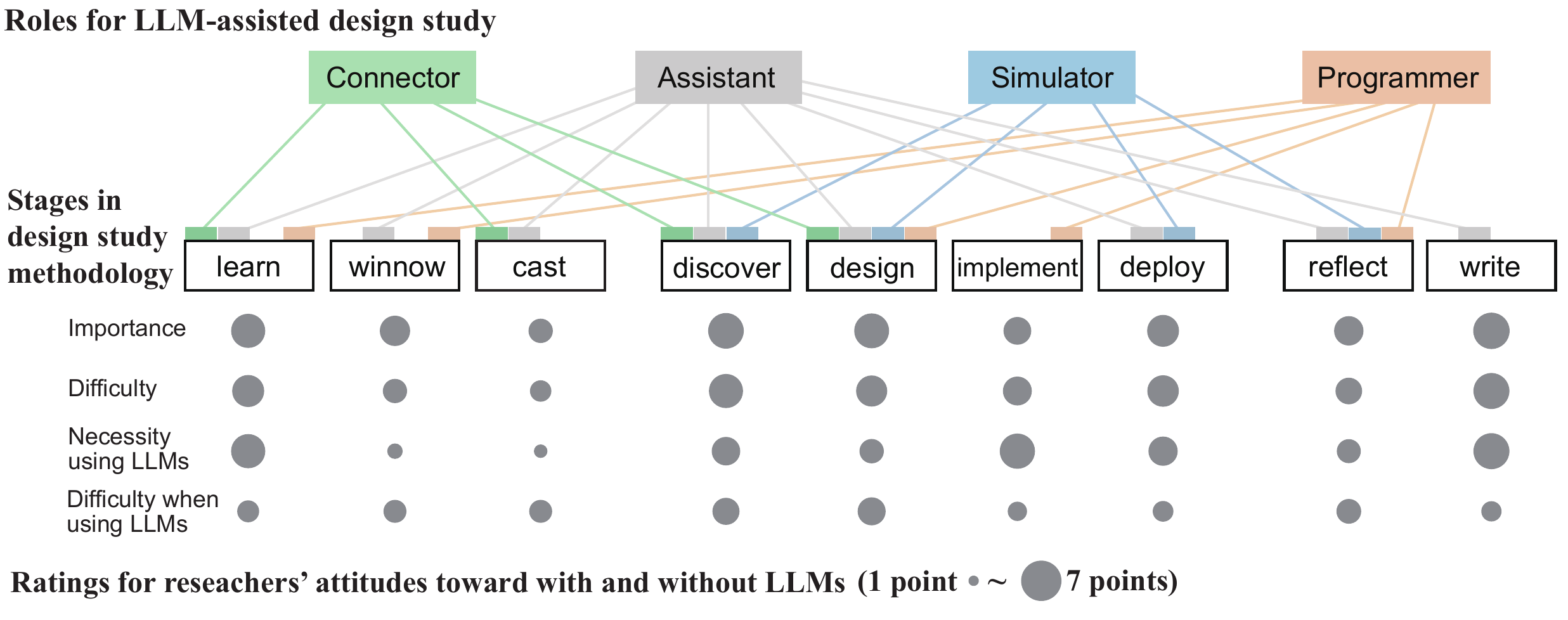}
  \caption{%
An overview of the summarized roles of LLMs in the LLM-assisted design study process. 
Following the framework proposed by Sedlmair et al.~\cite{sedlmair2012design}, we coded all nine stages of the design study by the compiled roles:
1) \textit{Connector:} LLMs are used to bridge gaps between visualization researchers and domain experts;
2) \textit{Assistant:} LLMs are commonly used for repetitive tasks to enhance productivity;
3) \textit{Simulator:} LLMs support simulating user behavior, predicting feedback, and testing workflows to uncover usability issues and hidden use cases;
4) \textit{Programmer:} LLMs are asked to assist in coding tasks, from prototyping to debugging.
The bottom section shows ratings from 30 researchers on the importance, difficulty, and necessity of using LLMs, as well as the difficulty when using LLMs at each stage.
}
  \label{fig:teaser}
}

\graphicspath{{figs/}{figures/}{pictures/}{images/}{./}} 

\usepackage{tabu}                      
\usepackage{booktabs}                  
\usepackage{lipsum}                    
\usepackage{mwe}                       

\usepackage{tikz}
\usepackage{xcolor}
\usepackage{ifthen}
\usepackage{etoolbox}

\usepackage{mathptmx}                  

\usepackage{graphicx}

\newcommand{\wen}[1]{\textcolor{black}{#1}}

\definecolor{mycolor1}{RGB}{22, 134, 191}    
\definecolor{mycolor2}{RGB}{234, 116,8}   
\definecolor{mycolor3}{RGB}{128, 0, 128}   

\newcounter{mysubsection}[section]
\renewcommand{\themysubsection}{\thesection.\arabic{mysubsection}}

\NewDocumentCommand{\mysubsection}{O{defaultsubseccolor} o m}{%
    \refstepcounter{mysubsection}%
    \vskip1ex\noindent%
    \begin{tikzpicture}[overlay,every node/.style={font=\large\bfseries\sffamily}]
        \node[right,minimum height=3.25ex,draw=#1,fill=#1,text=white,text width=5ex,align=center](section)at(0,0){%
            \IfNoValueTF{#2}{%
                \themysubsection
            }{%
                #2
            }%
        };
        \node[right,text=#1](title)at(section.east){#3};
        \draw[very thick,#1](section.north west)--([xshift=\linewidth]section.north west);
    \end{tikzpicture}
    \par\nobreak\vskip3ex
    \addcontentsline{toc}{subsection}{%
        \protect\numberline{%
            \IfNoValueTF{#2}{%
                \themysubsection
            }{%
                #2
            }%
        }#3%
    }%
}

\usepackage{tcolorbox}

\definecolor{color1}{RGB}{194, 234, 207}

\newcommand{\connector}{
\colorbox{color1}{connector}}

\definecolor{color2}{RGB}{178, 206, 232}

\newcommand{\simulator}{%
\colorbox{color2}{simulator}}

\definecolor{color3}{RGB}{247, 220, 185}

\newcommand{\programmer}{%
\colorbox{color3}{programmer}}

\definecolor{color4}{RGB}{229, 229, 229}

\newcommand{\assistant}{%
\colorbox{color4}{assistant}}

\newcommand{\revise}[1]{\textcolor{black}{#1}}

\begin{document}



\firstsection{Introduction}

\maketitle
Sedlmair \textit{et al.} introduced the term \textit{design study} to describe an applied research methodology that focuses on creating visualizations to address specific, real-world problems \cite{sedlmair2012design}. 
More specifically, they defined a visualization design study as ``a project in which visualization researchers analyze a specific problem faced by domain experts and design a visualization system that supports solving this problem''~\cite{sedlmair2012design}.
Building upon this definition, a nine-stage methodology framework was proposed for conducting design studies, as shown in Fig.~\ref{fig:teaser}, which has become a common guidance for this field of research in the visualization community. 
Design studies are crucial because they bridge the gap between theoretical visualization research and practical applications, ensuring that visualization tools are both usable and useful in real-world contexts \cite{meyer2019criteria}. 
Basically, researchers conduct design studies through labor-intensive processes (\textit{e.g.}, connecting with domain experts, extracting requirements via interviews, and iteratively designing and developing visualization systems),
which require significant effort and refined knowledge to gain effective solutions.

Meanwhile, Large Language Models (LLMs) have shown increasing popularity in recent years due to their impressive problem-solving capabilities with natural language~\cite{zhao2023survey}.
Thus, LLMs are now widely used to enhance various aspects of the design study process, including
assisting in creative problem-solving, 
handling large volumes of data,
and generating insightful findings~\cite{kim2024phenoflow,zhao2024leva,liu2024smartboard,shin2024visualizationary}.
By leveraging these advantages, LLMs show the potential to significantly streamline the workflow of design studies and make its process more efficient and effective~\cite{hutchinson2024llm,basole2024generative}.

Despite the increasing use of LLMs in conducting design studies, 
the question of how LLMs can effectively assist researchers in this domain remains underexplored. 
For example, researchers with limited expertise need to ``broaden'' their practice of LLM-assisted design studies to reach higher levels of proficiency, while those with a moderate or higher level of expertise can grasp the experience from peers to quickly ``deepen'' their strategies.
However, existing research
~\cite{shen2024data,he2023wordart,hou2024c2ideas} on how LLMs can aid co-design methodologies in other fields~\cite{rozo2025prompt, shen2025dashchat, gatti2024chatgpt,chiarello2024generative,swanson2024virtual,sun2024llm,kim2024understanding} cannot be easily generalized to the field of visualization due to its visualization-specific workflows and methodologies. 
To this end, a systematic study on how to facilitate the LLM-assisted design study process is urgently needed in the visualization community,
which points out several specific challenges for achieving an effective study procedure.
First, design study researchers often utilize LLMs in the way on their own understanding and personal experience, without any systematic and rigorous stage-by-stage guidance as a reference. 
Second, researchers are often unaware of the potential challenges they may face during the study, which can lead to significant confusion and require time-consuming efforts for correction~\cite{walny2019data}.
Even though these challenges have been recognized, 
\wen{individual researchers often lack an effective approach to address them.}
\wen{Third, the absence of a high-level summary of LLM characteristics makes it difficult for design study researchers, especially those with relatively less experience, to build an intuitive mental model of utilizing LLMs effectively.}

To fill the research gap, we conducted a multi-stage qualitative study that investigated common strategies and challenges for design study researchers using LLMs.
To provide a holistic perspective, we invited 30 design study developers and researchers from five different countries with varying levels of visualization expertise (\textit{i.e.}, novice, advanced beginner, intermediate, proficient, and expert) to share their insights.
More specifically, to address the first need above, we first conducted interviews with those participants to explore their \textbf{\textit{strategies}} for \wen{using LLMs for} conducting design studies. 
We then asked them follow-up questions to identify the \textbf{\textit{challenges}} they encountered while implementing these strategies and also collected the practices they used to address them.
To further characterize the types of tasks that LLMs can actually help with, we performed a post-study analysis and derived a set of \textbf{\textit{roles}} that LLMs may play during the design study stages by synthesizing all the above feedback. 
In addition to the interviews, we also incorporated a questionnaire to gather more nuanced insights into how researchers navigate each stage of the design study methodology~\cite{sedlmair2012design} via comparing their viewpoints with and without LLM assistance.
Finally, we distilled implications from all the findings that can inform both visualization practitioners and LLM researchers of how to achieve a more productive LLM-assisted design study process in the future.



\section{Related Work}

This section provides an overview of the existing literature relevant to this study, which can be categorized into three major areas: 
Human-LLM collaboration in visualization,
LLM-assisted co-design process, and 
qualitative study in visualization.

\subsection{Human-LLM Collaboration in Visualization}

Several studies have explored the dynamics of human-AI collaboration, particularly in the context of interactive machine learning and visual analytics.
Saha et al.~\cite{saha_humanAI_2022} examined the design and evolution of interactive machine-learning interfaces and pointed out usability challenges and interaction paradigms that optimize collaboration. 
Similarly, Kovalerchuk et al.~\cite{kovalerchuk_visual_knowledge_2021} investigated the role of AI in visual knowledge discovery and also identified key challenges for AI integration in visualization areas. 
Another relevant study by Schelble et al.~\cite{schelble_human_agent_teams_2020} analyzed shared mental models and trust in human-agent teams.
Furthermore, Chen et al.~\cite{chen_humanAI_coInnovation_2023} presented a case study on human-AI co-innovation, showcasing how AI augments human creativity and analytical tasks to enhance problem-solving capabilities.
Fill et al.~\cite{fill_visualization_AI_2023} provided an overview of how LLMs contribute to accessibility and efficiency in visualization tools. 
Researchers also developed an LLM-powered visualization system to support complex data analysis. 
Kim et al.~\cite{kim_visually_situated_NLU_2023} proposed a framework that improves LLMs' understanding of visual data to enhance user interactions with visualization systems.
Recent developments in natural language interfaces for visualization include Maddigan et al.~\cite{maddigan_chat2vis_2022}, which introduced a system that translates natural language queries into data visualizations. 
Similarly, Narechania et al.~\cite{narechania_nl4dv_2021} developed a toolkit that enables users to specify and translate them into visual representations to streamline analytics workflows. 
Similarly, Zhao et al.~\cite{zhao_lightva_2024} proposed LightVA, a lightweight visual analytics system leveraging LLMs for automated task planning and execution in data analysis workflows.
Kim et al.~\cite{kim2024chatgpt} studied how could ChatGPT be used in the educational scenario.
Additionally, Xu et al.~\cite{xu_LLM_lowLevel_tasks_2023} assessed the effectiveness of LLMs in performing low-level analytic tasks on SVG visualizations, demonstrating their potential for enhancing efficiency in data interpretation. 
Moreover, prior work explored leveraging LLMs for visual workflow generation  from the perspectives of financial Q\&A tasks~\cite{zeng2023flowmind}, industrial applications~\cite{koziolek2024llm}, and business process automation~\cite{sufi2022ai, xu2024llm4workflow}.

Despite some insightful knowledge being yielded from the above studies, they barely investigated the high-level summary of how LLMs can be used in the process of preparing, performing, and synthesizing design studies in visualization, which can directly inform design study researchers of a more effective usage of LLMs.

\subsection{LLM-assisted Co-design Process}

Apart from the visualization field, an increasing integration of Large Language Models (LLMs) across various domains has also led to extensive research for contextualizing the role of LLMs in assisting co-design workflows.
As an active field of utilizing LLMs, Human-Computer Interaction (HCI) researchers have explored its potential to enhance user experience, model transparency, and human-AI collaboration. 
Specifically, Kim et al.~\cite{kim2024understanding} investigated the co-design process of LLM-powered human-robot interactions, which highlights the ability of these models to facilitate seamless communication and adaptive interactions. 
Similarly, Liao et al.~\cite{liao2024designerly} examined the need for model transparency in AI-powered user experience design and emphasized the importance of interpretability for designers working with AI-generated suggestions. 
Sun et al.~\cite{sun2024llm} explored the role of LLMs in UI/UX design by demonstrating their capacity to advance the development of interactive systems. 
Also, Fill et al.~\cite{fill2023conceptual} discussed how the LLMs can be used to generate and interpret various types of conceptual models, \textit{e.g.}, Entity-Relationship (ER) diagram.
The work was followed by a study of a framework called Conceptual Model Augmented Generative Artificial Intelligence (CMAG)~\cite{fill2024cmag}, which used conceptual models to validate and enhance the outputs of generative AI models for improving their reliability and interpretability.
Beyond HCI, LLMs have also shown great potential in facilitating co-design processes in medical applications. 
For example, Swanson et al.~\cite{swanson2024virtual} introduced a virtual lab framework in which AI agents designed novel SARS-CoV-2 nanobodies. 
The use of LLMs in the engineering field has also been studied. 
Chiarello et al.~\cite{chiarello2024generative} analyzed the role of generative LLMs in engineering design by identifying key challenges associated with automating design processes. 
Similarly, Gomez et al.~\cite{gomez2024large} explored the application of LLMs in complex system design to facilitate knowledge transfer, automate documentation, and enhance problem-solving in engineering workflows. 
Furthermore, previous studies have also investigated the LLM-driven design authoring tool for various objectives, such as dashboard designs~\cite{shen2025dashchat}, programming~\cite{shanbhag2025tewen, zamfirescu2025beyond}, and video generation~\cite{rozo2025prompt, ma2025follow, li2025shots}.

While prior work has demonstrated the capacity of LLMs to assist in creative problem-solving, knowledge synthesis, and automated design, challenges such as model reliability and domain-specific adaptation in the area of visual analytics remain under-explored, which has yielded a strong need in the visualization community.

\subsection{Qualitative Study in Visualization}

The integration of qualitative methods, such as interviews, into visualization research has been explored across various domains,
which provides valuable insights for informing the design and application of visualization tools.
More specifically, in the domain of data representation, Hogan et al.~\cite{hogan2016elicitation} introduced the Elicitation Interview Technique to capture users' experiences with visual data, which provides a structured method to enable researchers to better understand how users interpret and interact with visualizations.
Also, Batch and Elmqvist~\cite{batch2018interactive} investigated the gap in interactive visualization during initial exploratory data analysis via qualitative interviews with data analysts.
After that, they proposed a methodological framework for highlighting how pair interviews enhance data collection processes, particularly in collaborative visualization studies~\cite{akbaba2023pair}.
Meanwhile, Wu et al.~\cite{wu2022defence} also initiated a workshop to discuss the critics of the contribution and rigor of visual analytics.
In the context of public health in recent years, Çay et al.~\cite{cay2020covid} explored user experiences with COVID-19 maps using remote elicitation interviews, revealing how users interpret pandemic-related visualizations and identified challenges in designing effective visual communication tools.
Visualization-specific user perspectives have also been explored in statistical and technical domains. For instance, the role of visualization in inferential statistics was studied in \cite{statisticians2024visualization}. 
Similarly, challenges and opportunities in distributed tracing visualization were characterized by Davidson et al.~\cite{tracing2024qualitative}.


However, the prior studies rarely explored how emerging tools, like LLMs, could help facilitate the design study process in visualization. 
This paper addresses this gap by investigating the potential strategies and challenges of utilizing LLMs to streamline workflows and foster more efficient and creative approaches during the design study process.
\section{methodology}

To explore the strategies, challenges, and roles of the LLM-assisted design study process, we conducted a multi-stage qualitative study, including participant profiling, an interview, a follow-up questionnaire, and a post-study analysis.
This section introduced our methodological approach.
We first described the participants involved, the data collection methods employed, 
and lastly, the procedures followed throughout the study.

\subsection{Participants}

\begin{figure}[t]
  \centering 
  \includegraphics[width=0.95\columnwidth
  ]{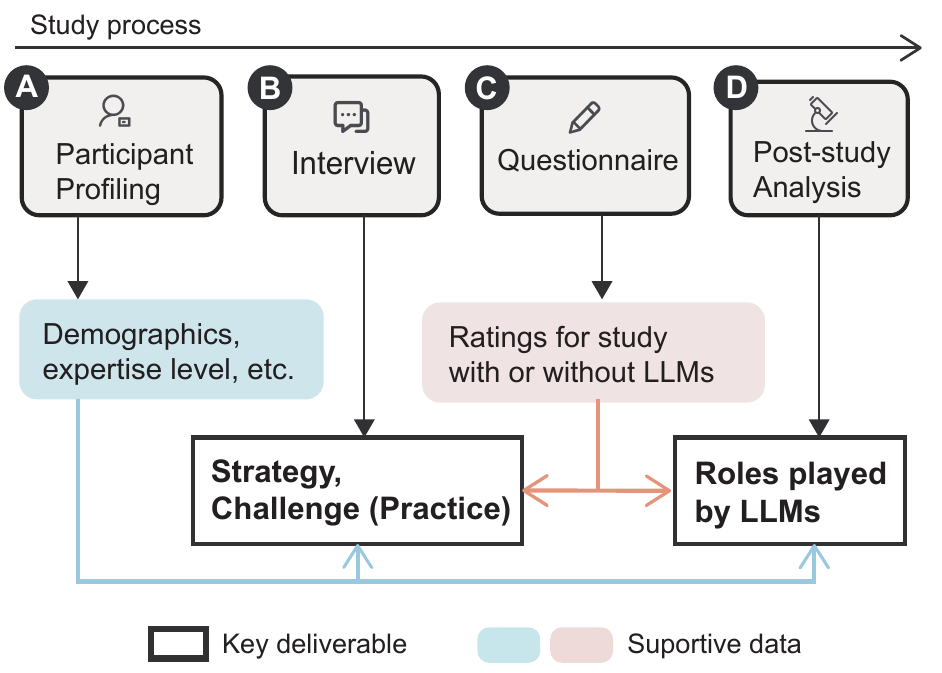}
  \caption{
 Our study consists of four steps: (A) participant profiling, (B) interview process, (C) questionnaire, and (D) post-study analysis. The outcomes of supportive data from (A) and (C) are used to support the in-depth analysis of interviews (B) and post-study analysis (D).
  }
  \label{fig:1}
\end{figure}

\begin{figure*}[t]
  \centering 
  \includegraphics[width=\linewidth
  ]{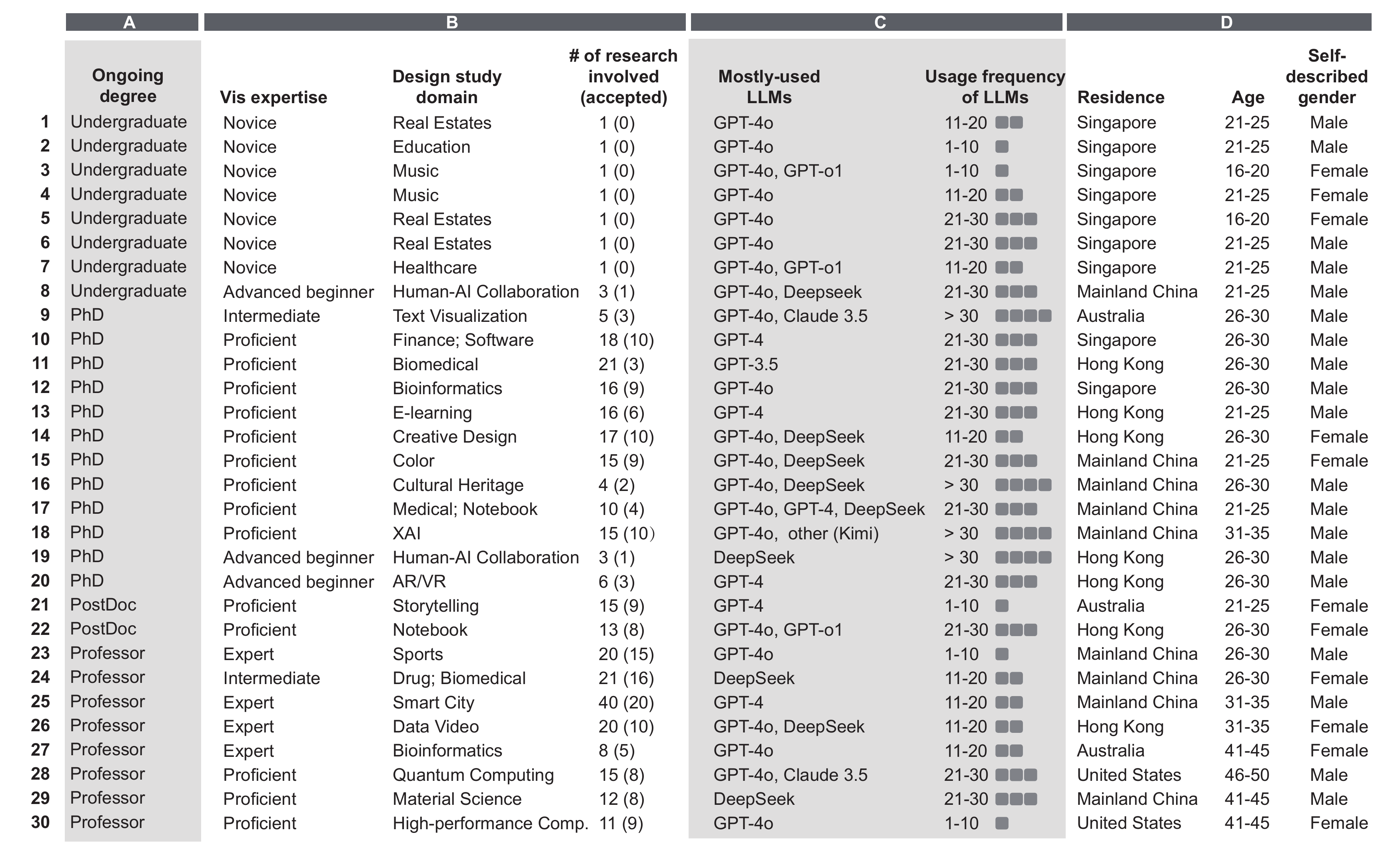}
  \caption{
\revise{
The profile information of 30 participants in our study. All dimensions are grouped into four categories: educational background (Area A), visualization study engagement (Area B), LLM utilization (Area C), and demographics (Area D). 
We sorted all participants according to their ongoing degrees.
All participants' names are anonymized.
}
  }
  \label{fig:2}
\end{figure*}

We recruited 30 paid participants via direct email request, with the compensation of 15 SGD in the form of
a gift card.  
As listed in Fig.~\ref{fig:2}, the participants were affiliated with 11 institutions located across Singapore, mainland China, Hong Kong, Australia, and the United States. 
Their academic professional roles included professors, Ph.D. candidates, post-doctoral researchers, and undergraduate students.
With this participant group, we aimed to survey a diverse range of perspectives about how LLMs are perceived and utilized across varying levels of experience.

Participants’ expertise in visualization and design study was identified by themselves into five levels: novice, advanced beginner, intermediate, proficient, and expert. 
Meanwhile, the participants' design study research spanned a wide range of interdisciplinary domains, including biomedical science, finance, smart cities, sports, etc. 
In terms of research productivity, our participants had contributed to over 300 research papers in total, of which 179 were accepted in peer-reviewed venues. 
Novice participants typically had limited publication records, often having contributed to one study.

Furthermore, we also requested participants to report the most-used LLM variants and the frequency of LLM usage in their workflows. 


\subsection{Data Collection}

\revise{We illustrate our data collection process in Fig.~\ref{fig:1},
as we planned to use the visual representation to better illustrate the study workflow.}
All interviews were conducted via online video calls.
Before each interview, we asked if the participant agreed with the recording, 
and we recorded the whole session upon their consent. 
At the same time, we took notes and created sketches based on the visualizations described by the participants during each session.
We shared these notes and sketches with participants to make sure they accurately reflected participants' original intentions. 
Additionally, we documented general impressions and recalled details immediately to capture any points that may not have been noted in real-time.

For the stages of participant profiling (Fig.~\ref{fig:1}(A)) and questionnaire (Fig.~\ref{fig:1}(C)),
we verbally inquired of all interviewees during the meeting right before and after the qualitative interview.
Note that all participants consented to provide their demographic information before we moved on to the stage of participant profiling.
All data was collected via Google Sheets and every participants can access and edit it anytime.

\subsection{Procedure}


\textbf{Participant profiling.}
Before the interviews began, participants were asked to provide demographic information,
their visualization expertise and LLM usage frequency. 
Additionally, participants were asked to indicate their primary design study domain.
In addition to the investigation of their demographics, educational background, and visualization study engagement illustrated above, we also asked participants to report their most-used LLM variants and the frequency of usage to characterize their reliance on LLMs. 
The LLM usage frequency refers to the approximate number of prompts used per day during they conduct the design study process.
We categorized the number into four ranges: 0–10 prompts per day, 11–20 prompts per day, 21–30 prompts per day, and more than 30 prompts per day. 
The participant profiling stage was brief, with each session lasting approximately 5 minutes.

\begin{figure}[t]
  \centering 
  \includegraphics[width=0.95\columnwidth
  ]{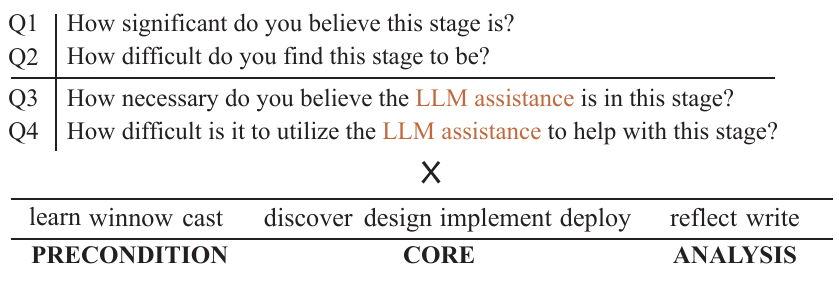}
  \caption{
The questionnaire questions upon the interview. 
All participants were invited to rate each question regarding the nine stages of the framework of the design study.
  }
  \label{fig:3}
\end{figure}

\textbf{Interview.}
At the beginning of each interview session, we first introduced the Sedlmair et al.~\cite{sedlmair2012design}'s nine-stage framework to participants. 
To ensure clarity, we provided detailed explanations of each stage, supplemented by examples to help participants understand the purpose of each stage. 
After that, we conducted one-on-one, semistructured, hour-long interviews with each participant. 
The interviews were designed around the three phases outlined in the original paper (as illustrated in Fig.~\ref{fig:3}), \textit{i.e.}, 
precondition, core, and analysis. 
During the interviews, we first asked participants to describe the strategies they typically employ for each phase in their routine tasks. 
Subsequently, we discussed with participants about the challenges or pitfalls in applying these strategies and how they attempted to address these challenges.

\textbf{Questionnaire.}
After completing the interviews, all participants were asked to fill in a post-study questionnaire. 
This questionnaire included four questions for each individual stage in the nine-stage framework, totaling 36 items (Fig.~\ref{fig:3}).
Specifically, participants were first asked to rate the \textit{importance} and \textit{difficulty} of each stage without LLMs. 
Then they also rated the \textit{necessity} of each stage requiring the assistance of LLMs,
followed by the rating for asking how \textit{difficult} it is to utilize LLMs in this stage.
These ratings were based on a 7-point Likert scale, where 1 indicated ``not important/difficult at all'' and 7 indicated ``extremely important/difficult''. 
The purpose of the questionnaire was twofold:  
1) To understand how much participants value each stage and perceive its difficulty when performing the design study process without LLM assistance; 
2) To explore the perceived necessity of using LLMs for each stage, as well as the difficulty of incorporating LLM assistance into the process.
We used the responses to this questionnaire to complement the qualitative feedback gathered during the interviews,
making it more insightful for summarizing the strategies and compiling the roles in the subsequent phases.

\textbf{Post-study analysis.}
We conducted a post-interview analysis to identify the high-level roles that LLMs can play in the design study process. 
The goal was to unify the diverse strategies employed by participants with different backgrounds and experiences.
We leveraged the information from participant profiling and their questionnaire responses to more accurately characterize individual researchers. 
To compile the roles, a team of three researchers independently reviewed the collected qualitative data, including interview transcripts, questionnaire responses, and participant profiles. 
Each researcher proposed an initial set of roles based on their interpretation of the data. These roles were then discussed in a series of group meetings to reach a consensus. 
In cases where disagreements arose, the team revisited the original data and engaged in iterative discussions to refine the roles until all members reached an agreement. 

\section{Roles of LLMs in Design Studies}

In this section, we introduce the roles of LLMs in supporting the design study process.
By analyzing the strategies collected from the 30 visualization researchers during the qualitative interviews, we identified four roles, \textit{i.e.}, \connector, \simulator, \programmer, and \assistant.
We first explain the roles briefly and then present the main strategies related to these roles in detail in Section \ref{sec:4}.

\textit{Connector}:
The \connector{} role focuses on bridging the gap between visualization researchers and domain experts.
More specifically, LLMs in this role help researchers from the visualization community learn domain-specific knowledge, help to extract design requirements, and work with domain experts more effectively.
\wen{For novice researchers, LLMs basically serve as a learning tool that provides basic background knowledge about the application domain, helping them quickly understand specific concepts.}
For visualization researchers, on the other hand, LLMs assist in analyzing domain-specific workflows, identifying key tasks, and collecting practical design requirements.
For example, a proficient PhD student could use an LLM to \wen{record and summarize} how bioinformatics researchers analyze tumor tissue slices, ensuring their visualization aligns with their real-world tasks. 
Experts might use LLMs to update their knowledge of a certain domain or generate drafts for academic papers based on high-level summaries.
Additionally, LLMs support communication by helping researchers articulate their needs to domain experts and translate domain-specific terminology into plain language.

\textit{Simulator}: 
For the \simulator{} role, LLMs are often used to simulate responses to questionnaires, predict potential user feedback, or test outcomes of workflows. By simulating user behavior, LLMs help researchers uncover hidden use cases, evaluate system usability, and even identify cognitive load issues before presenting the system to real users. 
Additionally, LLMs can take on the role of a critical reviewer, challenging the reasoning and logic behind a system and suggesting potential improvements.
For instance, a proficient PhD student might use an LLM to refine their system by simulating user responses during the evaluation process before they actually present the system to target users for testing. 
This role is most frequently relied upon by proficient PhD students who are experienced enough to build simulated insights into their design studies effectively.

\textit{Programmer}:
The \programmer{} role is a commonly used applications of allowing LLMs to help generate, optimize, and revise code, streamlining the technical implementation process from prototyping to debugging.
Specifically, researchers at varying levels of expertise, including novices, intermediates, and proficient researchers, reported using LLMs for programming support.
For novices, LLMs provide guidance for setting up their code frameworks or generating simple scripts, such as creating mock datasets or pre-processing data. 
Intermediate researchers use LLMs for more complex coding tasks, such as creating JavaScript code by translating hand-drawn drafts into a simple visualization, while proficient researchers are able to leverage LLMs to simplify SVG elements or optimize a visualization’s runtime performance.

\textit{Assistant}:
The \assistant{} role \wen{refers to the general-purpose support LLMs offer to boost researchers' productivity by automating repetitive or time-consuming tasks. 
Common users include searching for academic papers, summarizing a collaborator's background from a CV, or generating scripts from user study recordings.
}
For example, a novice researcher might use an LLM to summarize a complex academic paper into key points they can easily understand. 
A postdoc researcher could use it to transcribe interview recordings and distill key takeaways from them. 
Unlike the other roles, this role is used by researchers across all experience levels. 


\section{Qualitative Interview Results}
\label{sec:4}

In this section, we present the qualitative feedback collected from 30 participants, focusing on their experiences with leveraging LLMs to assist design studies in individual stages of the Sedlmair et al. framework~\cite{sedlmair2012design}. 
We explore the \textit{strategies} they employed, the \textit{challenges} they encountered, and any useful \textit{practices} they adopted accordingly. 
To explicitly indicate clearly identify the roles introduced above, we assign labels to findings by placing the corresponding LLM role name adjacent to each strategy.
Our analysis aims to uncover diverse practical insights in detail that emerge across participants with varying levels of visualization expertise and LLM usage frequency. 

\vspace{5mm}  


\mysubsection[mycolor1][1]{Precondition: Learn}

\textit{
``In summary, LLMs were found to be highly effective in aiding novice researchers to quickly grasp domain fundamentals, while more advanced users leverage them to stay updated with recent developments, together with a ``validation loop'' between visualization researchers, domain experts, and LLMs. 
However, challenges such as inaccuracies in technical details and confirmation bias need users to take practices like consulting domain experts for reliability.''}

\vspace{2mm}

The ``Learn'' stage indicates the need to acquire a solid understanding of visualization literature and techniques to inform the design study effectively.
Most novice visualization developers found LLMs particularly helpful for quickly grasping fundamental concepts in a specific domain \connector{} that would otherwise require extensive manual searching.
However, proficient and advanced users use LLMs to acquire the latest developments in their fields in a short period of time.

Despite these benefits, participants actually faced challenges, particularly novices, who often treated LLMs as search engines without the expertise to discern inaccuracies in the responses. 
For example, P5 mentioned that ``LLMs occasionally provide incorrect answers, but as a beginner, it is almost impossible to tell if there are actually any incorrect contents.'' 
To overcome this uncertainty, novices are often forced to double-check information with people with more expertise with a significant cost of time.
Another common challenge was the accuracy of detailed explanations provided by LLMs. While their high-level summaries were generally correct, technical details were sometimes erroneous, as noted by P23, who gave an example where the explanation of an object detection algorithm was conceptually accurate, but the technical details regarding the model's trainable layer were incorrect.
To address these issues, participants adopted various practices, including consulting domain experts and cross-referencing LLM outputs with literature \connector{}.

Another interesting finding is from advanced users. 
They noted a ``validation loop'', involving a dynamic process where visualization researchers, domain experts, and LLMs cross-verify with each other \connector{} \assistant. 
For example, visualization researchers used LLMs to validate domain knowledge provided by domain practitioners, while domain experts reviewed LLM-generated designing methods with visualization researchers to ensure alignment with real practice.
Meanwhile, visualization users also validate the LLM-generated domain knowledge with domain practitioners.
P21 described this triangulation process as effective for ensuring reliability but sometimes introducing pitfalls, because LLMs often aligned with the phrasing embedded in prompts.
One example is that if the domain experts ask LLM ``Please help me to decide if the stacked bar charts can fit my data or not,'' and LLM will most likely provide you with the answers about why stacked bar charts ``can'' fit your data instead of critical answers.
To mitigate this, P21 recommended allowing LLMs to generate independent evidence and conclusions without bias.

Furthermore, participants also highlighted some unique strategies. 
P24 used LLMs to fill gaps in domain-specific terminology\connector{}, enabling them to bridge knowledge gaps more efficiently. However, P21 also warned against relying on LLMs for summarizing design requirements, as these summaries often failed to align with their domain expertise. 
Similarly, 
P9 used LLMs to create coding tutorials and background explanations for self-training in new domains \connector{} \programmer.
But he noted the limitations that LLMs could not effectively use many emerging but immature tools,
even when provided with its detailed tutorial.

\vspace{5mm}  
\mysubsection[mycolor1][2]{Precondition: Winnow}

\textit{
``In summary, LLMs provide significant support in evaluating visualization necessity, identifying collaborators, and generating mock datasets. 
Despite challenges like hallucinated references and privacy concerns, participants employed strategies such as using LLM-powered platforms and refining generated content to ensure reliability and privacy.''}

\vspace{2mm}

The ``Winnow'' stage is about selecting promising collaborations with domain experts by filtering out unsuitable partnerships early in the process. 
Specifically, 
novice students often used LLMs to evaluate whether visualization techniques could actually solve a given problem \assistant. 
They prompted LLMs to provide the most relevant non-visualization papers to check if there were existing automatic approaches or algorithms that could address the problem. Additionally, they explored whether visualization methods, even from different areas, could be generalized and further applied into their contexts. 
However, a major challenge arose due to LLM hallucinations, as popular models frequently recommended papers that did not exist. 
To overcome this, participants turned to commercial LLM-powered literature searching tools to conduct more reliable literature reviews and verify the authenticity of suggested resources.
PhD students noted that prior to the availability of LLMs, they primarily relied on personal connections or pre-existing knowledge of researchers in the domain for collaborator recruitment. 
With LLMs, however, they could directly request a list of relevant researchers working in the same field \assistant.

Subsequently, some advanced researchers utilized LLMs to generate recruitment emails for inviting collaborators from other domains \assistant. 
While this approach saved time, it introduced specific challenges. 
First, the emails were often too generic, lacking personalized details such as how their research aligned with the project. 
To address this, participants attached the recipient's CV or homepage link to the LLM prompt, enabling the model to generate more tailored emails.
Second, privacy concerns arose when describing research projects to LLMs, as the model could extract sensitive keywords or overly detailed descriptions, risking unintentional idea leakage. To mitigate this, participants often have to manually verify generated content before sharing.

P11 highlighted using LLMs to address delays in obtaining datasets by generating mock datasets or expanding historical data while awaiting real data \programmer. 
Once real data was available, LLMs helped quickly adapt prototypes. Additionally, P11 used LLMs to draft Institutional Review Board (IRB) applications or consent forms, saving time while ensuring compliance with research protocols \assistant.

\vspace{5mm}  
\mysubsection[mycolor1][3]{Precondition: Cast}

\textit{``Generally, LLMs facilitate role assignment and collaboration by analyzing collaborators' expertise, and also act as agents to bridge interdisciplinary gaps. 
However, limitations in assessing collaborators' personal traits and achieving granularity required additional input from principal investigators (PIs) or `gatekeepers' to finalize decisions.''}

\vspace{2mm}

The ``Cast'' stage focuses on clarifying the roles of collaborators, ensuring that key participants, such as gatekeepers, are identified before starting the project.
According to our interview, advanced beginners in visualization used LLMs to assign roles within a project by providing the LLMs with collaborators' CVs or brief descriptions of their expertise \assistant. 
However, participants noted challenges in achieving enough granularity. 
For example, it was difficult for LLMs to distinguish whether a collaborator with a visualization background would be better suited as a designer or a programmer. 
Furthermore, LLMs lacked the ability to assess personal traits, such as leadership or communication skills, making it challenging to identify appropriate candidates for roles like team managers or connectors, who are the key members responsible for coordinating all members. 
To address these limitations, participants often relied on the Principal Investigator (PI) or a designated ``gatekeeper'' to make the final decisions regarding role assignments.

Fresh PhD students faced additional difficulties in distinguishing between front-line analysts and translators within their teams. 
Thus, they usually used LLMs to analyze collaborators' published papers and professional backgrounds, allowing the model to suggest suitable candidates for these roles \connector. 
Beyond role assignment, LLMs can also be directly regarded as an ``agent'' to facilitate collaboration among researchers from different domains \connector. For example, LLMs helped by translating domain-specific terminology into language that visualization researchers could understand. 


\vspace{5mm}  
\mysubsection[mycolor2][\textbf{4}]{Core: Discover}

\textit{``LLMs are useful in uncovering or validating domain-specific requirements and summarizing literature, though challenges like hallucinated references and misinterpretation of visual design contributions need manual validation and domain expertise.''}

\vspace{2mm}

The ``Discover'' stage highlights the engagement with domain experts to characterize the domain problems by iteratively refining understanding through discussions.
Novice researchers often used LLMs to explore the literature review by asking LLMs to suggest papers or summarize the latest research trends \assistant{} \connector, which was particularly helpful for those unfamiliar with the domain. 
However, this strategy often introduced hallucinated references to novices, where LLMs usually provided nonexistent papers or incorrect paper summarization.
According to P21, who is a professor and lecturer in visualization, LLMs often fail to recognize the graphical visual design elements, but the students cannot totally identify this incorrectness.
Proficient PhD students and professors used LLMs to summarize the takeaways from the academic paper \assistant, \textit{e.g.}, key contributions, methodologies, and takeaways from visualization papers. 
However, participants commented that LLMs often misinterpret the contributions of visualization papers, particularly related to understanding visual elements. 
For instance, the models often failed to accurately understand visual marks and channels for either visualization idioms or novel designs. 
This limitation required significant additional manual effort to validate outputs.

Beyond literature discovery, participants also explored interesting ways to use LLMs to understand domain-specific requirements \connector. 
Instead of relying solely on interviews with domain experts, some researchers used LLMs to analyze routine tasks performed by professionals in their fields. 
For example, P12 described using Copilot embedded in an IDE to document how bioinformatics researchers analyze tumor tissue slices on their laptops, 
which capture the practical workflows that might otherwise be overlooked in traditional interviews. 
This approach can significantly reduce information loss and provide a more precise understanding of domain requirements.

Participants also expressed that they often use LLMs to double-check the collected design requirements from domain experts \connector.
However, LLMs' answers are usually too generic to help with the requirement validation.
This limitation was particularly evident when researchers leveraged LLMs to generate design requirements or propose novel solutions. 
As a result, researchers emphasized the importance of combining LLM outputs with their own expertise, using the models as tools to augment, rather than replace, their discovery process.
P14 also mentioned the usage of the generation of the questionnaire for the pilot study or formative study \assistant. Also, P14 used LLMs to generate the possible answers from the provided questionnaire \simulator, which is used to compare with the real answers from the participants to identify some unique and insightful findings.

\vspace{5mm}  
\mysubsection[mycolor2][\textbf{5}]{Core: Design}

\textit{``In general, LLMs support visualization design by bridging gaps between designers and domain experts, generating design alternatives, and optimizing usability. 
However, limitations in global design understanding and integration into complex frameworks require researchers to refine LLM outputs with domain experts.''}

\vspace{2mm}  

The ``Design'' stage focuses on creating data abstractions, visual encodings, and interaction mechanisms based on the shared understanding developed in the discovery phase.
Three professors 
highlighted the use of LLMs as ``translators'' during the design process to bridge the gap between visualization designers and domain experts \connector. 
In this case, LLMs were used to facilitate communication by summarizing iterative meetings, generating meeting notes, and translating domain-specific knowledge and terminology into practical insights for visualization designers. 
Some PhD students reported using LLMs to decompose domain requirements, obtained during the ``Discover'' stage, into sub-tasks and brainstorm design ideas based on these sub-tasks \assistant. 
For example, they would ask LLMs to search approaches used in prior visualization studies, such as those published in IEEE TVCG papers, for similar design requirements. 

An advanced usage of LLMs involved refining initial visualization designs by analyzing code or design structures \programmer. 
Participants described the strategy that feeding LLMs the HTML elements of their designs (\textit{e.g.}, including SVG elements) along with the design requirements for understanding the structure of the visualization.
The LLMs were then asked to simplify or optimize the code to improve usability. 
Additionally, researchers used hand-drawn drafts as input for LLMs to generate simple JavaScript code \programmer, 
which was then validated to determine if the initial design could reveal the expected data patterns.

For system-wide design tasks, such as determining the overall layout, interaction logic, and workflow between views, LLMs were used to provide high-level suggestions based on the screenshot together with the code \simulator. However, participants noted that LLMs often struggled with these tasks because LLMs lacked the ability to maintain a comprehensive understanding of the global system. 
This limitation required researchers to manually refine and integrate the LLM's outputs into a cohesive global design.

LLMs were also employed as sources of design inspiration \assistant. Some researchers provided a domain and asked the LLM to suggest straightforward visualization designs to inspire their work, while participants noted that the suggestions were often limited to those commonly-used approaches. 
To address this challenge, researchers came up with practices such as incorporating evaluation metrics (\textit{e.g.}, significance or novelty) to assess the quality of the proposed designs. 
Moreover, PhD designers used LLMs to search commercial tools online for existing solutions as inspiration for their designs \assistant.

\vspace{5mm}  
\mysubsection[mycolor2][\textbf{6}]{Core: Implement}

\textit{``Basically, LLMs are widely used for data processing, refining prototypes, and code generation. Despite challenges like incompatibility with complex frameworks and difficulty in fine-tuning designs, 
participants adopt best practices such as providing detailed context and writing explicit code comments to improve outcomes.''}

\vspace{2mm}

The ``Implement'' stage in the design study methodology explores rapidly prototyping visualization tools while remaining flexible to changes based on ongoing feedback.
According to our interview, this stage is the common scenario where visualization developers rely on LLMs. 
First of all,
for data processing tasks, novices and PhD students can use LLMs to generate code for data cleaning or other pre-processing tasks \programmer. 
Also, after the system prototype becomes mature, LLMs can be employed to generate large-scale datasets to test the system's scalability \programmer, which can support researchers to initially evaluate if the prototype can seamlessly fit the domain's real needs.

In terms of technical stack setup, according to the findings from the professors, LLMs can help beginners to generate basic but functional project architectures \programmer, such as a simple React.js and Flask framework that supports fundamental tasks like data requesting.
This approach allows developers to focus on design implementation while avoiding the time-consuming process of learning code frameworks.

For implementing the design, LLMs can assist in generating code for visual design implementation \programmer. 
While LLMs work well for basic chart designs when provided with a sketch or bitmap, they often fail to recognize hierarchical structures. 
For example, in the case of a stacked bar chart, LLMs may generate separate bars instead of a cohesive stacked structure. 
As a workaround, novices may turn to user-friendly tools like Tableau, but it introduced new issues of flexibility and interactivity, limiting the visualization to only simple analysis tasks. 
Another significant challenge arises when the generated JavaScript code is only suitable for static HTML pages but fails to integrate into component-based frameworks like React.js. 
This incompatibility comes from several issues: 
1) complex interaction logic makes the code difficult to adapt; 
2) conflicts such as class name duplication can cause system bugs; 
3) the generated design may not align with the overall size or color scheme of the system;
and 4) SVG elements may not be correctly grouped.
The best practices to address these challenges include providing detailed context about all components to hint to LLMs. 

Several challenges also exist in the process of fine-tuning prototype systems. 
For example, it is difficult for LLMs to modify specific annotated parts of a design draft, as they struggle to recognize which part of the design is being referenced. 
Also, if a user requests modifying the ``top left'' circle, LLMs lack spatial awareness and cannot identify the target. 
Moreover, some descriptions, such as making a color scheme ``look fancy'', are often misunderstood by LLMs. 
A possible practice to address these issues is to write detailed comments in the code that explicitly specify the modifications required.


\vspace{5mm}  
\mysubsection[mycolor2][\textbf{7}]{Core: Deploy}

\textit{``Interestingly, LLMs are used to simulate user behavior, analyze performance, and formulate potential cases in the evaluation process. 
Challenges like generic outputs and hallucinations are mitigated by combining LLM insights with human expertise and testing with multiple models for diverse perspectives.''}

\vspace{2mm}

The ``Deploy'' stage basically means releasing the visualization tool in a real-world context and gathering feedback from domain experts to assess its effectiveness.
Advanced students have reported using LLMs to simulate user behavior, allowing the model to ``pretend'' to be a user \simulator. 
This approach helps uncover potential use cases for the system, as well as identify alternative workflows that differ from conventional ones proposed by human experts. 
However, two significant challenges arise: 1) the generic background knowledge of LLMs can lead to superficial stories that lack depth;and 2) hallucinations by LLMs may result in incorrect stories based on non-existent system features.

To evaluate whether the system adapts to real-world constraints, such as performance bottlenecks with large datasets, LLMs can be used to analyze backend network logs \assistant. 
For example, using Chrome's network log data, LLMs can identify performance problems related to data requests between the backend data port, which can detect bottlenecks that may hinder the prototype's performance. 
Also, novices often use LLMs to test example workflows to determine whether the cognitive load is manageable before presenting the system to real users \simulator. 
On the other hand, experts have explored using different variants of LLMs to test the system's performance \simulator{} more comprehensively. 
For example, models like GPT-4o are suited for reasoning tasks, while Claude may excel in logic-oriented evaluations. 
These experiments allow developers to assess how the system performs from the perspectives of various users.

Moreover, proficient students have mentioned some useful ways to support user studies and feedback analysis by LLMs. 
Specifically, developers can transcribe user study recordings into text and use LLMs to summarize key takeaways and feedback \assistant. 
Similarly, LLMs can assist in generating questionnaires and designing tasks based on the provided system workflow \assistant.
However, challenges exist in these practices. 
For example, relying on LLMs to summarize feedback may result in conclusions that do not align with the developer's domain expertise. Additionally, when generating questionnaires based on prior papers, LLMs may over-rely on the provided materials, leading to repetitive questions.

\vspace{5mm}  
\mysubsection[mycolor3][\textbf{8}]{Analysis: Reflect}

\textit{``To sum up, LLMs support reflective practices by simulating reviewer feedback and analyzing usability data. 
However, limitations in specificity and understanding user experiences require researchers to provide domain-specific data and detailed prompts to enhance feedback quality.''}

\vspace{2mm}

The ``Reflect'' stage emphasizes the critical analysis of the design study's outcomes, and refining existing guidelines based on new insights.
Advanced researchers used LLMs to simulate the role of a reviewer, challenging their research projects based on the abstract and identifying potential limitations \simulator. 
While this approach provided an initial critique, participants noted that the feedback was often too generic, focusing on broad issues such as unclear motivation, scalability, or generalizability of the proposed design.
To address this, participants adopted two practices to make the feedback more specific and actionable. 
First, they fed the LLM their own published papers to help the model learn the typical way of limitations relevant to their research domain. 
Second, they provided the LLM with a more specific direction, such as asking it to consider potential solutions in terms of the effectiveness of the approach used for resolving the problem.

Frequent LLM users highlighted the difficulty of capturing user experiences verbally via the evaluation process, 
particularly when investigating the usability of tools. 
To address this, they used LLMs to analyze performance data and assess whether the tool was suitable for real-world users \assistant{} \programmer. 
For example, participants prompted LLMs to generate code that could track cursor events within a visualization system, 
such as the duration of cursor hover in a view, click frequency of the SVG elements, or movement trajectories, which can be used to evaluate the usability of each view, 
with longer hover times and higher repetitive clicking frequencies generally indicating lower usability.
P13 specifically mentioned using LLMs to generate reflection summaries based on user interview scripts \assistant. 
Specifically, researchers can ask LLM  to provide the key insights and provide with a concise summary of user feedback by analyzing the transcripts.

\vspace{5mm}  
\mysubsection[mycolor3][\textbf{9}]{Analysis: Write}

\textit{``Generally, LLMs facilitate academic writing by generating drafts and refining text, especially for novices. Challenges like inconsistent domain terminology and lack of contextual awareness are addressed by training LLMs on domain-specific papers and adopting multi-step writing processes.''}

\vspace{2mm}

The ``Write'' stage focuses on documenting the findings in a coherent paper, and emphasizing clear abstractions and contributions. 
Novice students often leverage LLMs to generate papers based on high-level summaries they provided for each sentence, typically written in their native (non-English) language for international students \assistant. 
While this approach lowered the language barrier,
it introduced several challenges. 
First, the generated text often lacked detail, insight, or unique understanding, as the LLM tended to directly translate the student’s input.
Second, when using their native language to create summaries, LLM frequently struggled to capture the key points of a sentence, resulting in the misunderstanding of the user's intention. 
Participants at intermediate, proficient, and expert levels took a more structured approach to writing. 
More specifically, they provided high-level summaries for entire sections, allowing the LLM to draft longer passages \assistant. 
These drafts were then manually fine-tuned to ensure accuracy and alignment with the overall paper. 
A common challenge across all expertise levels was the lack of contextual awareness for the whole paper, 
which led to inconsistent terminology and style compared to the rest of the paper. 
To address these issues, participants used a practice of feeding LLMs domain-specific papers to learn the appropriate writing style beforehand and refine the generated text for consistency accordingly.

Some frequent LLM users among visualization researchers pointed out the process of combining the two steps above into one \assistant. 
Specifically, they first used reasoning-capable LLMs, such as GPT-o1 or DeepSeek, to first generate a detailed outline for a section based on a high-level summary. 
These tools then expanded the outline into more detailed summaries iteratively, down to the level of individual sentences.
Finally, advanced LLMs, such as GPT-4o, were used to craft each sentence based on the detailed summaries above. This multi-step process allowed for greater coherence in the final output.

However, some participants, such as P26, expressed reservations about relying on LLMs for academic writing. 
They argued that academic writing is a generative task, requiring the creation of original content (like the leap from nothing to something) based on human-distilled knowledge---an area where LLMs are less effective. 
Instead, they viewed LLMs as being better suited for tasks like summarizing or reporting, where the model works with existing material to produce concise summaries. 
This perspective highlights that novice writers may benefit most from LLM-generated drafts, because they tend to use LLMs for tasks that align with the model’s strengths, such as summarization.

\section{Guideline Summarization}

We proposed a set of role-based guidelines for LLMs' usage in deign studies, which are as follows:

\textit{\underline{Connector:}}
As an connector, LLMs can assist researchers in understanding domain-specific terminology and foundational concepts (\textit{Learn}) and support collaboration by analyzing the expertise of potential collaborators to suggest suitable roles within a project (\textit{Cast}).
Also, LLMs can assist in summarizing complex domain-specific workflows and extracting design requirements (\textit{Discover}).
In this stage, LLMs can also be employed to verify collected design requirements, ensuring that they align with the practical challenges of the domain.
Finally, in the \textit{Design} stage, LLMs can facilitate communication between visualization designers and domain experts by translating domain-specific knowledge into actionable design insights (\textit{Design}).

\textit{\underline{Simulator:}}
LLMs can simulate user behavior to uncover hidden use cases, alternative workflows, and potential usage scenarios (\textit{Deploy}).
They can also generate potential responses to questionnaires, enabling researchers to compare these with real participant feedback to identify unique insights (\textit{Discover}). Additionally, LLMs can simulate reviewer feedback to identify potential limitations in research projects, such as unclear motivation or scalability issues (\textit{Reflect}). They may also provide high-level suggestions for system-wide design tasks like layout and interaction logic as the basis for researchers to further refine and align with domain requirements (\textit{Design}).

\textit{\underline{Programmer:}}
LLMs in the programmer role can assist researchers in generating tailored coding tutorials and learning new technical stacks, such as emerging tools, to support self-training in new domains (\textit{Learn}). 
LLMs can generate mock datasets to address delays in obtaining real data and enable researchers to continue prototyping (\textit{Winnow}). 
Additionally, LLMs can create large-scale mock datasets to test system scalability and generate code for data cleaning and pre-processing tasks (\textit{Implement}). 
For visual design, LLMs can help with generating and optimizing code for prototypes and initial design(\textit{Design}, \textit{Implement}). 
Furthermore, LLMs refine specific parts of prototypes and generate usability tracking code, such as cursor events and click frequency, to assess system usability and layout effectiveness (\textit{Reflect}).

\textit{\underline{Assistant:}}
As the assistant, LLMs can aid researchers in filling gaps in domain-specific terminology and foundational concepts (\textit{Learn}) and support collaboration by evaluating visualization techniques or generating recruitment emails for collaborators (\textit{Winnow}). 
They can help summarize complex workflows, extract design requirements (\textit{Discover}). 
Additionally, LLMs can assist in translating domain-specific knowledge into actionable design insights, decomposing requirements into sub-tasks, and providing design inspiration (\textit{Design}). 
LLMs can analyze performance bottlenecks and summarize user feedback to evaluate system usability (\textit{Deploy}). 
Finally, they can summarize user interviews, and assist in drafting and refining academic papers (\textit{Reflect}, \textit{Write}).
\section{Implications}

In this section, we consider future directions for LLM-assisted design study 
and limitations of this work.

\subsection{Future Directions}
We distill from our study feedback some significant implications for the future of LLM-assisted design study.

\subsubsection{Toward a Project-Oriented LLM Copilot}

Current LLMs provide fragmented and localized assistance in design study projects, lacking the ability to capture the entire workflow holistically. 
As discussed in Section \ref{sec:4}, many of the challenges associated with LLM-assisted strategies stem from their inability to maintain an understanding of the project’s global context. 
For example, LLMs lose track of prior information when starting a new session and require users to repeatedly reintroduce the same content. 
Additionally, the narrative-driven prompts are often imprecise, making it difficult for the model to fully grasp the actual user intention. 
If possible, by integrating with various technical stacks---such as Overleaf for writing, Visual Studio Code for coding, or Figma for design---advanced LLMs could provide contextualized assistance through text writing, coding, or visualization design accordingly.
Hence, a future direction will probably focus on developing LLM systems that can naturally integrate into the entire design study workflow. 


\subsubsection{LLMs for Automated Task Decomposition}

Another promising direction for LLM-assisted design studies is the development of agent-based LLMs capable of autonomously decomposing coarse-grained tasks. 
Instead of requiring users to manually break down tasks or provide detailed prompts, 
these advanced LLMs could operate based on high-level commands, automatically decomposing them into sub-tasks and executing them seamlessly.
For example, given a set of collected design requirements, 
an agent-based LLM could independently perform task abstraction, 
and then it can generate the outputs, such as code from global visual representations (\textit{e.g.}, views) to detailed visual encodings (\textit{e.g.}, glyphs) in each view.

Furthermore, inspired by other domains, like chemistry, where multi-agent systems can already carry out autonomous research workflows. 
These systems can perform tasks such as hypothesis generation, experiment design, data collection, and analysis without requiring constant human oversight. 
However, compared to hard sciences like chemistry, visualization is inherently a ``soft science'' where human interpretation, creativity, and subjective judgment play a critical role. 
Thus, visualization often requires humans to remain in the loop to provide feedback.

\subsubsection{Are Human Researchers Indispensable in the Design Study Process?}

While LLMs have shown significant strength in assisting with various aspects of design studies, all participants commented that it is unlikely that they will fully replace human researchers in the foreseeable future. 
First, the process of brainstorming or assessing ideas cannot be entirely taken over by LLMs. 
Even though LLMs have a vast corpus of visualization knowledge, they struggle to determine the quality of an idea, for example, telling whether it is actually ``interesting'' or not.
Human researchers, on the other hand, leverage their subjective judgment, taste, and personal domain expertise to evaluate ideas based on higher-level abstractions built upon their accumulated knowledge in visualization. 

Second, while LLMs excel at summarizing existing knowledge, they fall short in creative thinking.
Although there have been advancements in enabling LLMs to generate creative outputs, the majority of their results remain generic and lack the originality required for groundbreaking insights. 
Tasks that need critical thinking or deep reflection are areas where LLMs are less capable, while human researchers can provide profound and insightful perspectives that are important to advancing the field.

Third, experts agreed that one of the most challenging parts of visualization design for LLMs is interaction. 
The success or failure of an interaction is highly dependent on the subjective experiences of domain users, which vary significantly across different systems and tasks. 
Evaluating and refining interactions requires iterative collaboration between visualization researchers and real users, 
thus it remains an open challenge for LLMs.

\subsection{Limitations}

While this study provides valuable insights into the roles and applications of LLMs in design studies, there are still some limitations.

\subsubsection{Trust and Usability Concerns}
Although most participants in our study expressed a positive attitude towards LLMs and their potential to support design studies, there are also concerns regarding trust and usability regarding LLMs. For example, some participants, such as P21, 
highlighted the ``dark side'' of LLM usage, where users may think of these tools as unreliable for specific tasks, 
such as fine-tuning the mining models integrated in an interactive system.
\revise{Second, although we have sought to include participants from diverse regions and research backgrounds, these participants may already have a predisposition toward using LLMs that do not represent the whole population of design study researchers.}
\revise{Third, while the paper proposes guidelines for LLM usage in design studies, it does not empirically validate these recommendations. Future work should involve assessing these guidelines in real-world scenarios.}

\subsubsection{Dynamic Progression of LLMs}
This paper focuses on the current generation of LLMs and how they are being used in design studies. 
However, it is obvious that LLMs are evolving rapidly, with continuous improvements in their capabilities and applications. 
Thus, some of the insights presented in this work may become outdated as more advanced LLMs emerge.  
Despite this limitation, we have provided deep discussions and practical guidelines for the design study field. 
For example, we provided potential research directions for better integrating LLMs into design studies in the future. 
Additionally, we have discussed why humans cannot be replaced by LLMs in the foreseeable future. 
These contributions provide value for future studies to build upon, even as LLM technology becomes increasingly advanced.

\subsubsection{Limited Focus on Domain- and LLM-Specific Usage}
This paper did not focus much on the detailed comparison of how LLMs are used across different domains or how various types of LLMs differ in their applications. 
For example, we did not conduct an in-depth comparison of how LLM usages are different between scientific domains (\textit{e.g.}, biomedical research or quantum computing) and application domains (\textit{e.g.}, smart cities or sports). 
\revise{Moreover, while the findings of this visualization-specific study can be potentially relevant to researchers and practitioners in other disciplines interested in human-AI collaboration (\textit{e.g.}, HCI), we need additional investigation to confirm and understand its broader generalizability.}
Similarly, we did not explore much about the differences in usage patterns between various types of LLMs (\textit{e.g.}, reasoning or coding).
Future research could address these gaps by conducting comparative studies in the above contexts. 

\section{conclusion}

This paper investigated the strategies, challenges, and roles associated with integrating Large Language Models (LLMs) into the visualization-specific design study process. 
Through a multi-stage qualitative study involving 30 participants across various expertise levels and domains, 
we identified four primary roles for LLMs: \textit{Connector}, \textit{Simulator}, \textit{Programmer}, and \textit{Assistant}. 
Each role supports researchers differently across the nine stages of the design study methodology.
Also, we revealed the attitudes towards how researchers treat LLM-assisted design studies regarding the importance and difficulty of using LLMs through the post-study analysis.
Moreover, we proposed specific useful guidelines to benefit all domain researchers with different expertise,
followed by a discussion of the potential implications for future research.

\acknowledgments{%
\revise{This project was funded in part by the Singapore Ministry of Education AcRF Tier 1 22-SIS-SMU-092 and Academic Research Fund Tier 2 (Proposal ID: T2EP20222-0049), and NTU Start Up Grant awarded to Yong Wang. 
}
}


\bibliographystyle{abbrv-doi-hyperref}

\bibliography{template}

\newpage
\clearpage

\appendix 

\section{Findings from the Ratings}

\begin{figure*}[h]
  \centering 
  \includegraphics[width=\linewidth
  ]{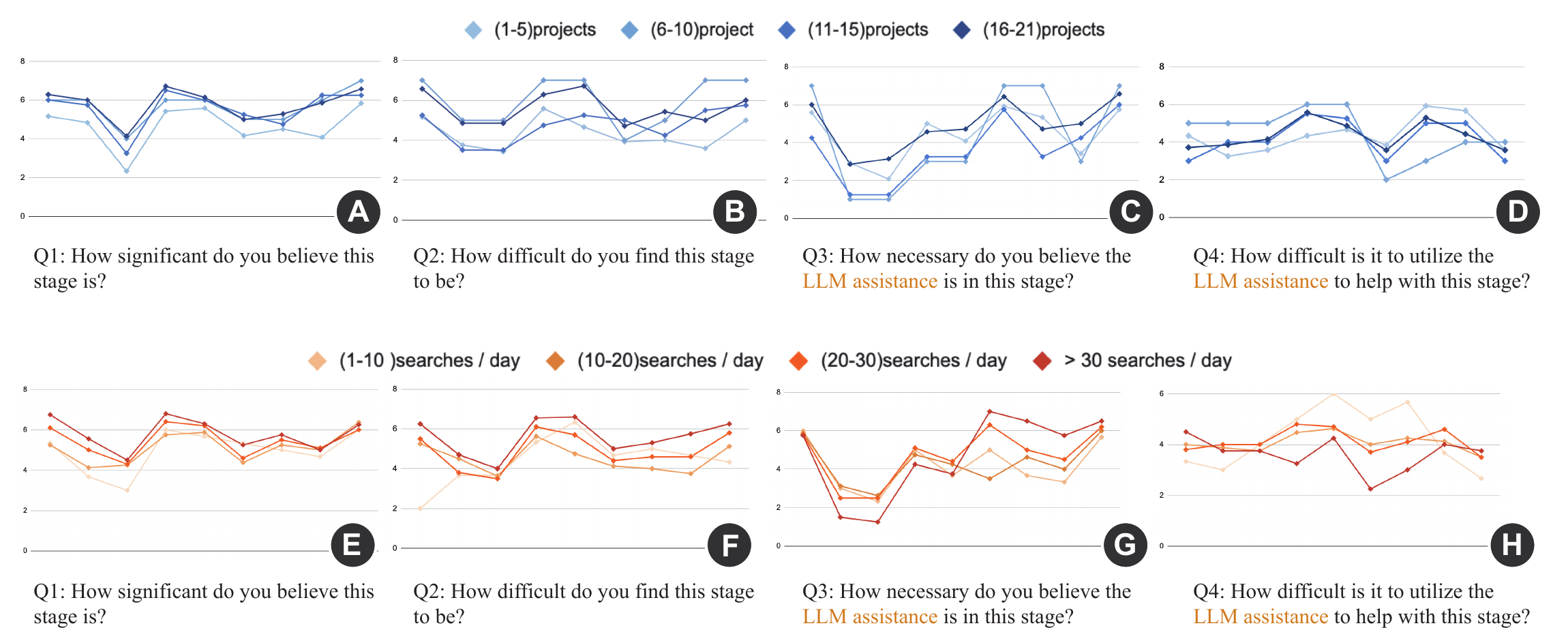}
  \caption{
The results of the questionnaire evaluate participants’ perceptions of the nine stages in the design study methodology. The first row of figures (A–D) categorizes participants by the number of research projects they have been involved in (1–5, 6–10, 11–15, and 16–21 projects), while the second row (E–H) groups them by their daily LLM usage frequency (1–10, 10–20, 20–30, and >30 searches/day). The x-axis represents the nine stages of the design study methodology, while the y-axis indicates the corresponding ratings on a 7-point scale. Each line series reflects the different participant groups based on project involvement or LLM usage.
}
  \label{fig:4}
\end{figure*}

From an overall perspective, the ratings of the four questions reveal distinct patterns across the nine stages of the design study methodology. 
For the first question (\textit{How significant do you believe this stage is?}), participants consistently identified the 1st stage (\textit{Learn}), 4th stage (\textit{Discover}), 5th stage (\textit{Design}), and 9th stage (\textit{Write}) as the most important stages, regardless of their visualization expertise or experience with LLM usage. Conversely, the 3rd stage (\textit{Cast}) and 6th stage (\textit{Implement}) were rated as the stages that mostly do not require LLM assistance, indicating a lower perceived significance for these stages in the absence of LLMs.
For the second question (\textit{How difficult do you find this stage to be?}), the 1st stage (\textit{Learn}), 4th stage (\textit{Discover}), 5th stage (\textit{Design}), and 9th stage (\textit{Write}) were consistently rated as the most challenging stages. In contrast, the 2nd stage (\textit{Winnow}) and 3rd stage (\textit{Cast}) were considered the simplest stages, reflecting lower difficulty across participants.
For the third question (\textit{How necessary do you believe the LLM assistance is in this stage?}), the ratings showed that with LLM assistance, the 1st stage (\textit{Learn}), 6th stage (\textit{Implement}), and 9th stage (\textit{Write}) were perceived as the stages most necessarily requiring LLM support. On the other hand, the 2nd stage (\textit{Winnow}) and 3rd stage (\textit{Cast}) were rated as the least necessary stages for LLM assistance.
For the fourth question (\textit{How difficult is it to utilize the LLM assistance to help with this stage?}), the findings indicate that the difficulty of all stages decreased compared to the scenario without LLM assistance. Notably, the coding-related stages, particularly the 6th stage (\textit{Implement}), were rated as the easiest to collaborate with LLMs, demonstrating the effectiveness of LLMs in reducing task complexity in these areas.

\subsection{Analysis for the First Question}

For the question \textit{``Q1: How significant do you believe this stage is?''},
we analyzed the ratings regarding visualization research expertise (measured by the number of involved papers), and it is evident that researchers with higher expertise in visualization research consistently provided larger ratings for each stage. 
This trend highlights their emphasis on the importance of each stage in the design study process. Notably, researchers with the lowest visualization expertise rated the reflection stage significantly lower compared to those with higher expertise. This suggests that novice students with limited visualization research backgrounds have yet to fully grasp the importance of reflection in visualization research.

For the ratings regarding LLM usage frequency (measured by the number of submitted prompts each day), a similar pattern emerges: researchers with higher usage frequency tend to provide larger ratings for each stage. 
However, the differences in ratings among researchers who use LLMs less frequently are much smaller in this case, as the ratings for each stage almost overlap with those of researchers who use LLMs more frequently. 
This indicates that the understanding of stage significance is less differentiated when considering LLM usage compared to their visualization expertise.

\subsection{Analysis for the Second Question}

For the second question (\textit{``How difficult do you find this stage to be?''}), the ratings provided by researchers with different levels of visualization expertise exhibit some distinct patterns compared to the first question. Specifically, the ratings for the stages before the sixth stage (\textit{Learn}, \textit{Winnow}, \textit{Cast}, \textit{Discover}, and \textit{Design}) are divided into two clusters. Researchers with the lowest and higher intermediate visualization expertise provided almost identical ratings, which are consistently lower than the ratings given by researchers with lower and highest visualization expertise. These patterns suggest that perceived difficulty in the first five stages does not vary linearly with expertise level. 
For the remaining four stages (\textit{Implement}, \textit{Deploy}, \textit{Reflect}, and \textit{Write}), there is no clear pattern in the ratings regarding visualization expertise. This suggests that the difficulty of these later stages is less influenced by visualization expertise and may be perceived similarly across all levels of expertise.

For the ratings based on LLM usage frequency, the same pattern emerges: researchers who use LLMs less frequently tend to rate the difficulty of each stage lower, indicating that they perceive the design study process to be generally simpler. 
Notably, researchers who use LLMs the least rated the first stage (\textit{Learn}) as significantly simpler than any other stage. 
This may be attributed to their limited understanding of the \textit{Learn} stage, leading them to underestimate the complexity of acquiring domain-specific background knowledge.

\subsection{Analysis for the Third Question}

For the question \textit{``How necessary do you believe the LLM assistance is in this stage?''},
We evaluated the researchers regarding visualization expertise, and the ratings reveal distinct patterns across the stages. 
Specifically, researchers with the highest and lowest visualization expertise consistently rated the first five stages (\textit{Learn}, \textit{Winnow}, \textit{Cast}, \textit{Discover}, and \textit{Design}) as more necessary for LLM assistance compared to those with intermediate expertise. 
However, for the last four stages (\textit{Implement}, \textit{Deploy}, \textit{Reflect}, and \textit{Write}), 
the ratings become more scattered, showing no clear trend.
Interestingly, novice researchers identified the 1st, 4th, 6th, 7th, and 9th stages (\textit{Learn}, \textit{Discover}, \textit{Implement}, \textit{Deploy}, and \textit{Write}) as the most necessary for LLM assistance. 
At the same time, they rated the reflection stage as the least necessary for LLM assistance, aligning with the findings from the first question, where novices underestimated the importance of reflection in visualization research.
On the other hand, the most experienced researchers rated the 3rd, 5th, and 8th stages (\textit{Cast}, \textit{Design}, and \textit{Deploy}) as the stages where LLM assistance is most needed, reflecting their nuanced understanding of where LLMs can provide the greatest value.

For the researchers regarding LLM usage frequency, the ratings for the first stage (\textit{Learn}) show consistent agreement across all levels of LLM usage, indicating a shared perception of its necessity for LLM assistance. 
For the 2nd to 5th stages (\textit{Winnow}, \textit{Cast}, \textit{Discover}, \textit{Design}), researchers with more frequent LLM usage rated these stages as less necessary for LLM assistance. 
This may be because experienced LLM users have recognized the limitations of LLMs in supporting these stages.
In contrast, for the last four stages (\textit{Implement}, \textit{Deploy}, \textit{Reflect}, and \textit{Write}), the pattern reverses: researchers with less frequent LLM usage rated these stages as less necessary, while frequent LLM users rated them higher. 
This trend likely reflects the experienced users' understanding of LLMs' strengths in these stages. Notably, for the 6th to 9th stages (\textit{Implement}, \textit{Deploy}, \textit{Reflect}, and \textit{Write}), the ratings are the highest among researchers who use LLMs frequently, demonstrating that LLMs can significantly assist with coding, deploying, and reflection tasks.

\subsection{Analysis for the Fourth Question}

For the fourth question (\textit{How difficult is it to utilize the LLM assistance to help with this stage?}), the ratings reveal distinct trends based on visualization expertise. 
For the first five stages (\textit{Learn}, \textit{Winnow}, \textit{Cast}, \textit{Discover}, and \textit{Design}), researchers with the lowest visualization expertise rated these stages as the easiest to interact with LLMs. 
In contrast, for the last four stages (\textit{Implement}, \textit{Deploy}, \textit{Reflect}, and \textit{Write}), they rated these as the most difficult to use LLMs effectively. 
Researchers with intermediate or high visualization expertise provided ratings that were consistently moderate, as indicated by their rating lines falling between the highest and lowest lines across all nine stages. 
This suggests that researchers with more experience perceive the difficulty of using LLMs as relatively balanced across the stages.

For the researchers regarding LLM usage frequency, the first three stages (\textit{Learn}, \textit{Winnow}, and \textit{Cast}) show similar ratings across all levels of LLM usage. Notably, researchers who rarely use LLMs rated the first two stages (\textit{Learn} and \textit{Winnow}) as the simplest to use LLMs, indicating that they find it straightforward to interact with LLMs for these initial tasks. 
After the third stage (\textit{Cast}), a clear pattern emerges: the more frequently researchers use LLMs, the easier they perceive the later stages to be. This is particularly evident for the coding and deployment stages (\textit{Implement} and \textit{Deploy}), where frequent LLM users rated the interaction with LLMs as very easy. 
Interestingly, for the last two stages (\textit{Reflect} and \textit{Write}), researchers who do not frequently use LLMs also rated these stages as relatively easy and straightforward to interact with LLMs. This suggests that even infrequent LLM users recognize the utility of LLMs in assisting with reflection and writing tasks.

The findings suggest that visualization expertise and LLM usage frequency significantly shape participants' perceptions of the design study stages. 
Novice researchers tend to undervalue certain stages, such as reflection, and overestimate the simplicity of early stages like \textit{Learn}. Frequent LLM users demonstrate a more nuanced understanding of LLM capabilities, recognizing its strengths in later stages like coding and deploying while acknowledging its limitations in earlier conceptual stages.
These insights highlight the need for tailored training and support for novice researchers to better understand critical stages like reflection and for optimizing LLM tools to address challenges in early-stage tasks. Furthermore, the results underline the importance of leveraging LLMs effectively for coding, deployment, and writing tasks, where their utility is most apparent.

\end{document}